\documentclass[aps,prl,amsmath,amssymb,groupedaddress,twocolumn]{revtex4}
\usepackage{amssymb}
\usepackage{graphicx}
\usepackage{color}

\begin{document}

\title{Observation of {\it empty liquids} and {\it equilibrium gels} in a colloidal clay}

\author{Barbara Ruzicka$^{1,*}$,  Emanuela Zaccarelli$^{2,*}$, Laura Zulian$^3$, Roberta Angelini$^1$, Michael Sztucki$^4$, Abdellatif Moussa\"{\i}d$^4$,
Theyencheri Narayanan$^4$, Francesco Sciortino$^2$}
\affiliation{$^1$ CNR-IPCF and Dipartimento di Fisica, Universit\`a di Roma La Sapienza, Piazzale A. Moro 2, I-00185, Rome, Italy.\\
$^{2}$ CNR-ISC and Dipartimento di Fisica, Universit\`a di Roma La Sapienza, Piazzale A. Moro 2, I-00185, Rome, Italy.\\
$^3$CNR-ISMAC via Bassini 15, 20133 Milan, Italy.\\
$^4$ European Synchrotron Radiation Facility, B.P. 220 F-38043
Grenoble, Cedex France.\\
$^*$e-mail: barbara.ruzicka@roma1.infn.it;
emanuela.zaccarelli@phys.uniroma1.it}

\maketitle

{\bf The relevance  of anisotropic interactions in colloidal
systems has recently emerged in the context of the rational design
of novel soft materials~\cite{Glotzer}. Patchy colloids of
different shapes, patterns and functionalities~\cite{reviewpatchy}
are considered the novel building blocks of a bottom-up approach
toward the realization of self-assembled  bulk materials with
pre-defined properties~\cite{pine,mowald, mirkin,kegel,gang}.
 The possibility of tuning the interaction anisotropy will make possible to recreate molecular structures at the
nano- and micro- scale  (a case with tremendous technological
applications), as well as to generate novel unconventional phases,
both ordered and disordered. Recent theoretical
studies~\cite{BianchiPRL} suggest that the phase-diagram of patchy
colloids can be significantly altered by limiting the particle
coordination number (i.e., valence). New concepts such as empty
liquids~\cite{BianchiPRL} -liquid states with vanishing density-
and equilibrium gels~\cite{ZaccarelliRev,BianchiPRL,kobsastry}
-arrested networks of bonded particles, which do not require an
underlying phase separation to form~\cite{LuNature}- have been
formulated. Yet no experimental evidence of  these predictions has
been provided. Here we report the first observation of empty
liquids and equilibrium gels in a complex colloidal clay, and
support the experimental findings with numerical simulations. }

We investigate dilute suspensions of Laponite, an industrial
synthetic clay made of nanometer-sized discotic platelets with
inhomogeneous charge distribution and directional
interactions. Similarly to other colloidal
clays~\cite{Rennie,Lekkerkerker,Shalkevich}, Laponite has technological
applications in
 cleansers, surface coatings, ceramic glazes, personal care and cosmetic
products, including shampoos and
sunscreens~\cite{CumminsJNCS2007}. The anisotropy of the face-rim
charge interactions combined with the discotic shape of Laponite
produce a very rich phase diagram including disordered (gels and
glasses) and ordered (nematic) phases, on varying colloidal volume
fraction, at fixed ionic
strength~\cite{Mourchid_Lang_1995,Mongondry,RuzickaPRL,JabbariPRL2007,CumminsJNCS2007,Shahin_Lang_2010}.
At low concentrations the system ages very slowly up to a final
non-ergodic state~\cite{RuzickaPRL,JabbariPRL2007}.

In this work we extend the observation time to time-scales
significantly longer than those previously studied and discover
that, despite samples appear to be arrested on the second
timescale~\cite{RuzickaPRL,JabbariPRL2007}, a significant
evolution takes place on the year timescale. Samples undergo an
extremely slow, but clear phase separation process into clay-rich
and clay-poor phases that are the colloidal analog of gas-liquid
phase separation. Spectacularly the phase separation terminates at
a finite {\it but very low} clay concentration, above which the
samples remain in a homogeneous arrested state. At variance
with respect to the structural transition previously observed for
isotropic systems upon a variation of density of
depletants~\cite{Dibble} this thermodynamic phase transition is
driven by a change in colloid density (Laponite concentration).
The observed features are instead strikingly similar to those
predicted in simple models of patchy particles~\cite{BianchiPRL},
suggesting that Laponite forms an (arrested) empty liquid at very
low concentrations. Furthermore, differently from gels
generated by depletion interactions~\cite{piazza,LuNature}
 or from molecular glass-formers~\cite{sastryprl}, where arrest occurs after the phase
 separation process has generated high-density fluctuation regions, here  phase separation
 takes place in a sample which is already a gel.

\begin{figure*}[t]
\centering
\includegraphics[width=.6\textwidth]{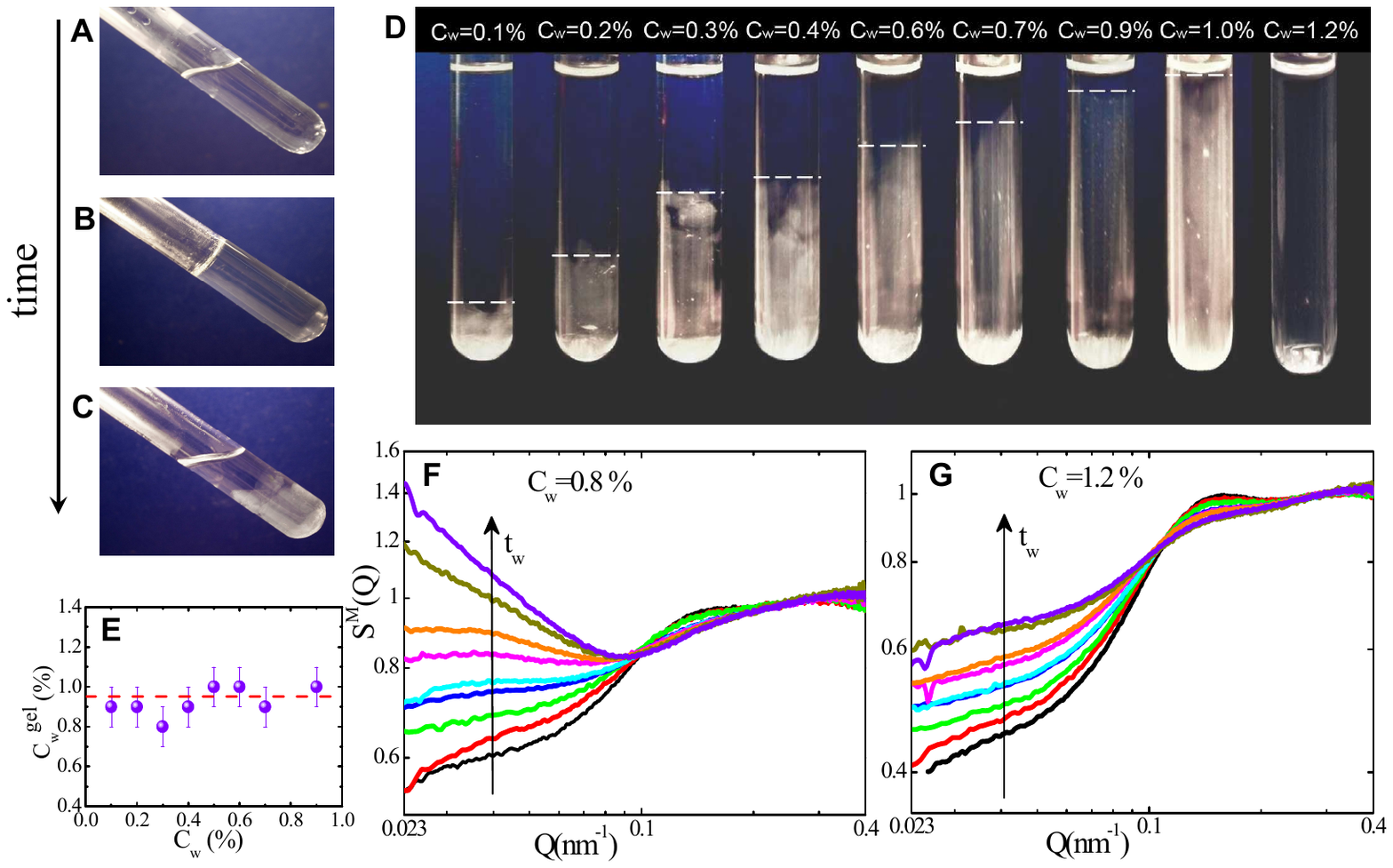}
\caption{{\bf Experimental behavior of diluted Laponite
suspensions.} {\bf a-c}, Photographs of  a $C_w= 0.4 \%$ sample
{\bf a}, in the initial fluid phase ($t_w=0$), {\bf b}, in the gel
state ($t_w\simeq4000$ h), {\bf c}, in the phase-separated state
($t_w\approx30000$ h). {\bf d}, Photographs of samples in the
concentration range $0.1 \le C_w \le 1.2 \%$ at very long waiting
times (about 30000 h). All samples with $C_w \leq 1.0 \%$ show a
clear evidence of two coexisting phases, separated by an interface
whose height  (dashed horizontal lines) increases progressively
with $C_w$. {\bf e}, Estimated concentration of the denser gel
phase in the
 separated samples shown in panel d. {\bf f}, Evolution of
the measured $S^M(Q)$ with waiting time for a $C_w=0.8 \%$ sample
(located inside the phase separation region). {\bf g}, Evolution
of $S^M(Q)$ with waiting time for a $C_w=1.2 \%$ sample (located
outside the phase separation region). The curves in panels {\bf f}
and {\bf g} are measured at increasing waiting times. From bottom
to top: $t_w$= 500, 900, 1600, 2700, 3400, 4700, 6000, 8700, 11000
h.} \label{FigDiagExp}
\end{figure*}

Fig.~\ref{FigDiagExp}a-c shows photographs of the temporal
evolution of a low concentration Laponite sample (weight
concentration $C_w$=0.4\%). The initially fluid suspension
(waiting time $t_w$=0) (Fig.~\ref{FigDiagExp}a) progressively
ages, forming a gel (the sample does not flow if turned upside
down as evident from Fig.~\ref{FigDiagExp}b). The gelation time,
as probed by dynamic light scattering, depends on clay
concentration and it is of the order of few thousand hours for low
concentration samples~\cite{RuzickaPRL}. Waiting significantly
longer time (several years), the sample undergoes a phase
separation, creating a sharp interface between an upper
transparent fluid  and a lower opaque gel
(Fig.~\ref{FigDiagExp}c). Phase separation is observed for all
samples with  $C_w \lesssim 1.0 \%$. Fig.~\ref{FigDiagExp}d shows
a photograph of different concentration  samples about three years
after their preparation. The height of the colloid-rich part
(indicated by the dashed lines in Fig.~\ref{FigDiagExp}d)
increases progressively with $C_w$, filling up the whole sample
when  $C_w \approx 1.0 \%$. This value thus marks the threshold of
the phase separation region. We note also that the denser phase in
all samples with $C_w <$ 1.0 \% approaches $C_{w}^{\ \ gel}
\approx 1.0 \%$, i.e. exactly the coexisting liquid density, as
reported in Fig.~\ref{FigDiagExp}e. This dense phase retains a
memory of the separation process, remaining turbid even at very
long times. The turbidity  is due to the formation of large
density fluctuations, generated during the phase separation
process, whose length scales are comparable to the ones of visible
light. Instead, higher concentration samples ($C_w
>$ 1.0 \%) do not show any phase separation and maintain their
arrested and transparent character at all times (see $C_w=1.2 \% $
sample in Fig.~\ref{FigDiagExp}d). Furthermore, no macroscopic
changes are observed, in the entire concentration range, in the
following four years (seven years, around 60000 hours, in total).

\begin{figure*}[t]
\centering
\includegraphics[width=.65\textwidth]{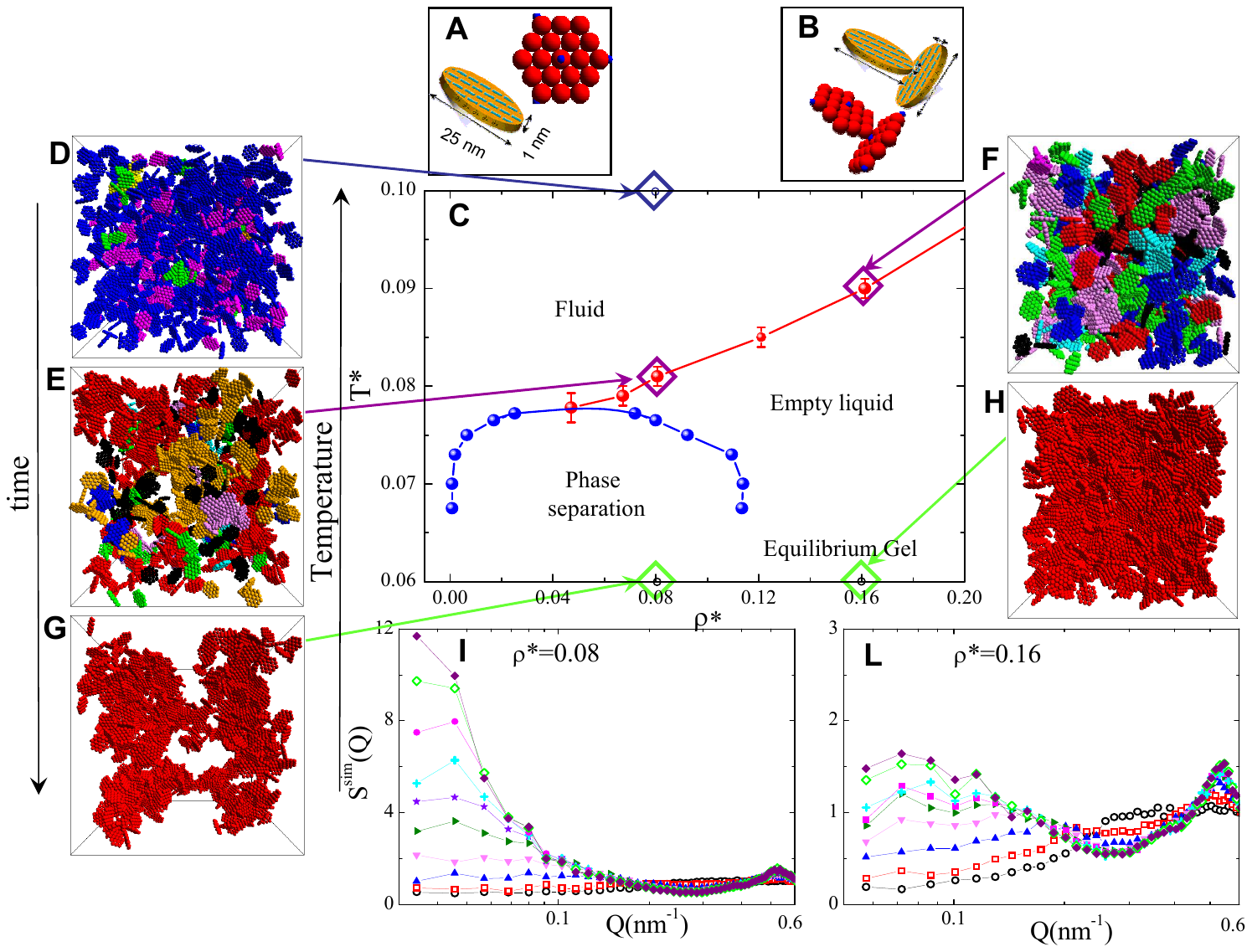}
\caption{{\bf Behavior of the patchy particle model for Laponite
discs.} {\bf a}, Cartoon of a Laponite platelet and its
schematization as a rigid disc composed by 19 sites (red spheres)
with 5 attractive patches (blue spheres), three located on the rim
and one  at the center of each face. {\bf b}, Cartoon representing
a $T$-bonded configuration for two interacting Laponite platelets
and its realization in simulations. {\bf c}, Numerical phase
diagram: binodal (blue curve) and percolation  locus (red curve)
in the $\rho^*-T^*$ plane, where $\rho^*$ is the number density
scaled by the close-packing density and $T^*$ is the thermal
energy scaled by the strength of the bond (see Methods). {\bf
d-h}, 3D snapshots of MC simulations at different state points.
Different colors correspond to different clusters, and the
red-color is reserved for the percolating cluster: {\bf d},
equilibrium fluid phase at $T^*=0.10$ and $\rho^*\simeq 0.08$;
{\bf e,f}, equilibrium configuration at percolation for
$\rho^*\simeq 0.08$ and $0.16$; {\bf g,h}, final gel
configurations at $T^*=0.06$ inside ($\rho^* \simeq 0.08$) and
outside ($\rho^* \simeq 0.16$) the phase separation region. In
these cases, platelets are connected into a single cluster (gel),
which is clearly inhomogeneous (homogeneous) inside (outside) the
binodal region. {\bf i,l},~Evolution of the $S^{sim}(Q)$ after a
quench  at $T^*=0.06$  for  $\rho^*\simeq 0.08$,  (inside the
phase separation region) and  $\rho^* \simeq 0.16$    (outside the
phase separation region). Waiting times are: $10^2, 1.2 \times
10^5, 5.7 \times 10^5, 1.6 \times 10^6,  3.6 \times 10^6, 6.1
\times 10^6, 10^7, 2.2 \times 10^7, 4.9 \times 10^7, 1.1 \times
10^8$ in MC steps.} \label{FigDiagSim}
\end{figure*}

To connect the absence of phase separation and of turbidity
in samples above the coexisting liquid density with the proposed
equilibrium gel concept~\cite{BianchiPRL}, as well as to characterize the
structural evolution of the phase separation, we have 
investigated the evolution of the structure of Laponite samples
for more than one year. Through Small Angle X-ray Scattering
(SAXS) measurements, we have monitored the static structure factor
$S^M(Q)$ at different waiting times, from the initial fluid phase
up to the gel state (arrested according to light scattering
measurements) and during the initial stages of the phase
separation process. We have focused on two different
concentrations, respectively inside ($C_w=0.8 \%$) and outside
($C_w=1.2 \%$) the phase separation region. The measured $S^M(Q)$
are reported in Fig.~\ref{FigDiagExp}f and g for different waiting
times $t_w$. At short times $S^M(Q)$ shows a peak, induced by the
overall electrostatic repulsion, at $Q \sim$0.15 nm$^{-1}$, i.e. a
distance of $\approx40$ nm which corresponds to platelets
considerably far away one from each other. With increasing $t_w$,
this peak disappears in favor of a new peak emerging at $Q
\gtrsim$ 0.4 nm$^{-1}$, which corresponds to roughly contacting
platelets in a $T$-configuration ($\lesssim15$ nm). The shift of
the main peak to higher $Q$ values is accompanied by a progressive
increase of the intensity at small wave vectors, indicating the
onset of aggregation for both concentrations. However with the
proceeding of the aging dynamics a drastically different behavior
for the two samples is observed: while at $C_w=0.8 \%$ the small
$Q$ intensity continuously increases (Fig.~\ref{FigDiagExp}f),  at
$C_w=1.2 \%$ the intensity saturates to a constant value at late
times (Fig.~\ref{FigDiagExp}g), as also shown in Fig.~S1 in
Supplementary Information.

The continuous increase of the small $Q$ behavior of $S^M(Q)$ for
$C_w < $ 1.0 \% signals the ongoing phase-separation process,
revealed also by the sample turbidity.  More interesting is the
interrupted growth of $S^M(Q)$, accompanied by the formation of a macroscopic gel, which is observed for
 $C_w > $ 1.0 \%.  Since gelation must occur via a percolation transition,
 the system has organized itself into a spanning network. The saturation in the evolution of  $S^M(Q)$,
 as time proceeds further,  indicates that  the system has reached its long-time equilibrium structure,
 i.e. a stable network. The absence of further structural changes is consistent with the low effective valence of
platelets, which at this point have formed most of their possible bonds. Due to the low density of the system,
the final gel state is rather non-compact, as signaled by the finite value of $S^M(Q)$  at small $Q$.
During the entire aggregation process the system remains always transparent, confirming that
fluctuations on a length scale comparable to the wavelength of light do not develop.

These results suggest that the samples outside the phase
separation region reach their equilibrium structure on the
year-time scale, while the samples inside the unstable region,
despite their apparent gel state, slowly evolve toward complete
phase separation. Thus, in Laponite, gelation precedes, but does
not preempt, phase separation.

\begin{figure*}[t]
\centering
\includegraphics[width=.65\textwidth]{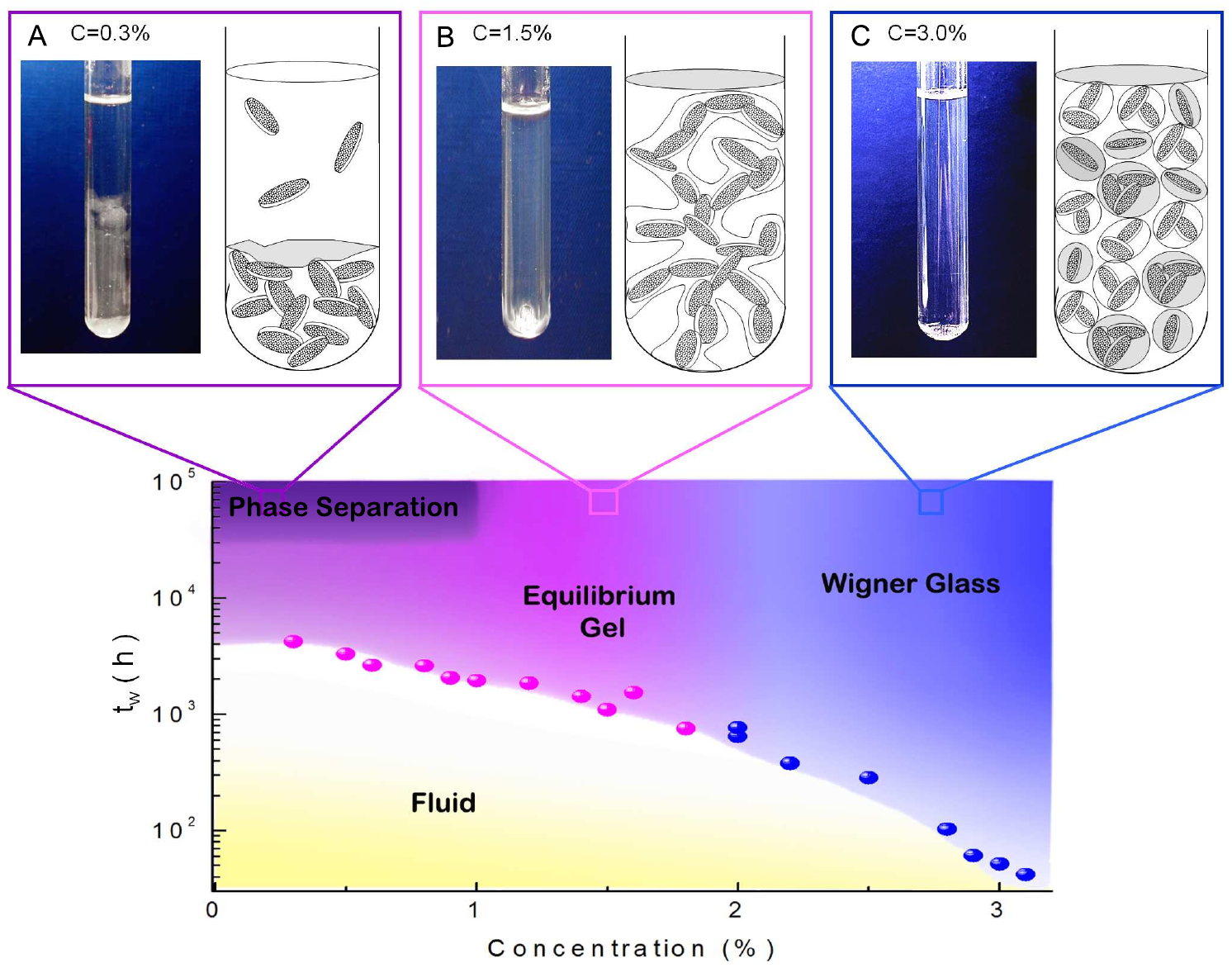}
\caption{{\bf Phase diagram  of diluted Laponite suspensions, in
the waiting time vs. concentration plane, resulting from the
combined experimental and numerical results.}   Symbols correspond
to experimental $t_w$ values requested  to observe non ergodic
behavior according to DLS~\cite{RuzickaPRL}; boundaries inside
colored regions are guides to the eye. For long waiting times,
three different regions are identified, whose representative
macroscopic behavior and a pictorial microscopic view are reported
in {\bf a-c}, .
 {\bf a}, Phase-separated sample with colloid-poor (upper
part) and colloid-rich (lower part) regions for $C_w \leq 1.0 \%$.
{\bf b}, Equilibrium gel  for $1.0 < C_w <2.0 \%$, characterized
by a spanning network of $T$-bonded discs.
 {\bf c}, Wigner glass, expected for  $2.0 \leq C_w  \leq 3.0 \%$~\cite{newprl},  where disconnected platelets
are stabilized in a glass structure by the electrostatic repulsion, progressively hampering the formation of $T$-bonds.
}\label{FigDiagLapo}
 \end{figure*}

To favor the interpretation of the previous results we introduce a
primitive model of patchy discs, which aims to mimic the strong
rim-face charge attraction and the tendency of Laponite clay to
form open structures. Each platelet is schematized as a hard disk,
following the work of Ref.~\cite{HansenSimu}. To implement the
rim-face linking ($T$-bonds~\cite{djikstra,odriozola})  between
different platelets, each disc is decorated with three sites on
the rim and one at the center of each face (five sites in total).
Only face-rim bonds can form, and they are modeled with a
short-range square-well attraction, ensuring that each site can be
involved at most in one $T$-bond. A representation of the model is
provided in Fig.~\ref{FigDiagSim}a and b.

We perform Monte Carlo (MC) and Gibbs ensemble MC (GEMC)
simulations to evaluate the gas-liquid coexistence region in the
reduced density $\rho^*$- reduced temperature $T^*$ plane.
Fig.~\ref{FigDiagSim}c shows the binodal line,  i.e. the locus of
points  separating homogeneous and phase-separated state, and the
percolation line, defined as the line separating a finite cluster
fluid phase (Fig.~\ref{FigDiagSim}d) from configurations
characterized by the presence of a spanning infinite (transient)
cluster. Fig.~\ref{FigDiagSim}e and f show snapshots of the
simulated system at the percolation line. Consistently with
previously studied patchy-spheres models~\cite{BianchiPRL}, the
gas-liquid coexistence region is confined in a narrow window of
$T^*$ and $\rho^*$. Indeed, the coexisting liquid density, scaled
by the closed packed value, occurs at $\rho^* \approx 0.114$, in
significantly dilute conditions. Hence, a wide region of densities
exists, above the coexisting liquid density, where the system can
be cooled down to very low $T$ without encountering a phase
separation, giving rise to an empty liquid
state~\cite{BianchiPRL}. Such state is composed by an
extensively-bonded percolating network that, at low $T$,
restructures itself on a timescale which exceeds the observation
time, generating an equilibrium gel state.

We also study the out-of-equilibrium dynamics of the model.  To
mimic the experimental protocol, we first equilibrate the system
at high $T$ (corresponding to sample preparation) and then
instantaneously quench it (corresponding to $t_w=0$) to  a
sufficiently low $T$, so that the bond-energy is large as compared
to the thermal energy as in Laponite~\cite{Mourchid_Lang_1995}. We
propose to interpret the experimental behavior as the low-$T$
limit of our model, connecting the increasing  waiting time  to a
progressive temperature decrease in the  numerical
study~\cite{SoftMatterSciortino}. Snapshots of the final
configurations of the system, after a quench inside and outside
the phase coexistence region, are shown in Fig.~\ref{FigDiagSim}g
and h. Independently from the density of the quench, the final
configuration is always characterized by a single spanning cluster
incorporating all particles. The structure of such cluster is
highly inhomogeneous for quenches inside the coexistence region
(Fig.~\ref{FigDiagSim}g) and homogeneous for quenches in the empty
liquid region (Fig.~\ref{FigDiagSim}h). Since at these low $T$ the
bond lifetime becomes much longer than the simulation time, the
bonded network is persistent, i.e. the system forms a gel.

Fig.~\ref{FigDiagSim}i,l show the static structure factors
calculated from MC configurations $S^{sim}(Q)$ at several times
(in MC-steps, equivalent to $t_w$)  following the quench, for two
densities, respectively inside and outside the phase separation
region. On increasing waiting time the scattered intensity
increases in the region of the contact peak ($T$-bonds, $Q
\approx$ 0.5 nm$^{-1}$), the peak that monitors the aggregation
kinetics, revealing the bond formation process. The notable
feature in $S^{sim}(Q)$ is the increase of the scattering at small
wavevectors. As in the experimental data, two different scenarios
occur at long times after preparation, respectively  for samples
inside (Fig.~\ref{FigDiagSim}i) and outside
(Fig.~\ref{FigDiagSim}l) the unstable region. While inside the
phase separation region $S^{sim}(Q)$ at small $Q$  increases
indefinitely,  outside this region the growth stops after a finite
waiting time, showing no further evolution.

The zero-th order model introduced here for describing Laponite at
low densities condenses the electrostatic interactions between
opposite charges  into short-ranged attractive sites and neglects
the overall repulsive electrostatic interactions, in the spirit of
primitive models~\cite{nezbeda}. These simplifications lead to a
different shape of $S(Q)$ at short times -- controlled in Laponite
by the screened electrostatic interactions~\cite{newprl}(see
Supplementary Information)
--- and in the absolute values of $S(Q)$ at small $Q$.
However, the qualitative features shown by the model (Fig. 2i,l)
coincide with the ones measured experimentally (Fig. 1f,g) both
inside and outside the unstable region, pointing out that our
model correctly captures the essential ingredients for describing
the experimental behavior. Most importantly, indefinite increase
(saturation) in the growth of  $S(Q)$ at small $Q$ is seen only
inside (outside) the region where phase separation is observed,
both in experiments and in simulations.

The present results suggest a novel phase diagram of Laponite
suspensions, summarized in Fig.~\ref{FigDiagLapo}, including the
crossover taking place for $C_w \sim 2.0\%$ toward a Wigner
glass~\cite{newprl}. At low concentrations, for $C_w \lesssim
1.0\%$, the system evolves via a sequence (Fig. 1a-c) of
clustering (hours-days), gelation (months)  and phase separation
(years), from a  sol to a homogeneous gel to a phase separated
sample in which only the dense phase is arrested. This progression
with $t_w$ is strongly reminiscent of a constant-density path in
the equilibrium phase diagram in which temperature is
progressively decreased and the system evolves from a sol
(Fig.~\ref{FigDiagExp}a, Fig.~\ref{FigDiagSim}d), to a percolating
structure (Fig.~\ref{FigDiagExp}b, Fig.~\ref{FigDiagSim}e),
finally encountering the phase-separation region
(Fig.~\ref{FigDiagExp}c, Fig.~\ref{FigDiagSim}g). Differently from
the case of isotropic short-range attractive
colloids~\cite{LuNature} where a homogeneous fluid is driven by a
spinodal decomposition into an arrested network, here the system
first forms a gel and then the gel extremely slowly increases its
local density  to fulfill the search for  a global free energy
minimum (phase separated) state. These features are exactly the
ones predicted to take place in patchy colloidal systems when the
average valence is small~\cite{SoftMatterSciortino},  strongly
supporting the view that the observed  phase-separation is a
genuine effect of the directional interactions.  The gel phase
observed in Laponite above $C_w=1.0\%$ can thus be interpreted as
an arrested empty liquid state, generated by the reduced valence,
spontaneously arising from the combination of  the platelet shape
and the patchy distribution of opposite charges on the disc
surface. The observed phase separation  on the year time-scale
calls attention on the fact that the  long term stability of soft
materials is controlled by the underlying phase diagram. Knowledge
of thermodynamic properties is thus crucial in designing material
with desired properties. Our case study shows that  a careful
choice of the density (within the empty liquid region) may provide
materials which are extremely stable in the long term (gels that
do not phase separate nor age in the present case), since they are
formed continuously from the liquid state, but finally reaching
--- through a very slow dynamics  --- their equilibrium
configuration.

{\bf Methods}

\maketitle

{\bf Laponite sample preparation} Laponite RD suspensions were
prepared in a glove box under N$_{2}$ flux and were always kept in
safe atmosphere to avoid samples
degradation~\cite{ThompsonJCIS1991}. The powder, manufactured by
Rockwood Ltd, was firstly dried in an oven at $T$=400 C for 4
hours and it was then dispersed in pure deionized water
($C_s\simeq 10^{-4}$ M), stirred vigorously for 30 minutes and
filtered soon after through 0.45 $\mu m$ pore size Millipore
filters. The same identical protocol has been strictly followed
for the preparation of each sample, fundamental condition to
obtain reliable and reproducible results~\cite{CumminsJNCS2007}.
At the end of this preparation process, Laponite  forms a
colloidal dispersion of charged disk-like particles, with a
diameter of $\sim$25 nm and a thickness of $\sim$1 nm with both
negative charges  on the faces  and positive ones on the rims. The
distinct rim and face charges induce a directional face-rim
attraction, while a limited valence is realized by means of the
additional electrostatic repulsion between like-charges, which
inhibits the formation of a large number of bonds per particle.
The starting aging time (t$_w$=0) is defined as the time when the
suspension is filtered.

The estimate of the denser gel phase concentration $C_{w}^{\ \
gel}$ in the phase-separated samples is provided by the ratio
between the nominal concentration and the volume occupied by the
dense phase, i.e. neglecting the gas phase concentration.

{\bf Small Angle X-ray Scattering} Small Angle X-ray Scattering
(SAXS) measurements were performed at the High Brilliance beam
line (ID2) at the European Synchrotron Radiation Facility (ESRF)
in Grenoble, France, using a 10 m pinhole SAXS instrument. The
incident x-ray energy was fixed at 12.6 keV. The form factor
$F(Q)$ was measured using a flow-through capillary cell. SAXS data
were normalized and the scattering background of water was
subtracted. The measured structure factor has been obtained as
$S^{M}(Q)=I(Q)/F(Q)$.

{\bf Simulations} Each platelet is modeled as a hard rigid disk
composed of 19 sites on a hexagonal mesh, inspired by the work of
~\cite{HansenSimu}. Each site is a hard-sphere of diameter
$\sigma$, as schematically shown in Fig. 2a. A comparison with
Laponite fixes $\sigma=5$ nm.   Each platelet is decorated with
five sites, three located symmetrically on the rim and  two on the
two opposite faces of the central hard-sphere. This primitive
model highlights the anisotropic nature of the platelet-platelet
interaction~\cite{HansenSimu,djikstra,odriozola} but neglects
 the repulsive electrostatic barriers which control
the timescale of the aggregation kinetics in Laponite (for a more
detailed discussion, see Supplementary Information).

Site-site interactions (acting only between rim and face sites)
are modeled as square well interactions, with range $0.1197\sigma$
and depth $u_0=1$.  The number of sites  controls the effective
valence of the model.  Since only rim-face bonds can be formed,
the lowest energy state is characterized by an average number of
bonds per particle equal to four. The exact choice of the valence
controls the location of the gas-liquid unstable region, but does
not affect the topology of the phase diagram. Indeed, we have
verified that upon variation of the number of rim charges, this
topology is preserved, in full agreement with the case of spheres
decorated by patches~\cite{BianchiPRL}.

Reduced temperature $T^*$ is  the thermal energy scaled by the
strength of the bond, $T^*=k_B T/u_0$, where $k_B$ is the
Boltzmann constant. Reduced density $\rho^*$ is defined as the
number density $\rho=N/L^3$, where $N$ is the number of particle
and $L$ the side of the cubic box, scaled by the closed packed
density, corresponding to a hexagonal close packing of discs
(which is space-filling and equal to $\sqrt(2)/19 \sigma^{-3}$).
Gibbs Ensemble MonteCarlo (GEMC) simulations are carried out for a
system of 250 platelets which partition themselves into two boxes
whose total volume is 66603$ \sigma^3$, corresponding to an
average number density $\rho^*\approx 0.05$. At the lowest studied
$T$ this corresponds to roughly 235 particles in the liquid box
and 15 particles in the gas box (of side 32 $ \sigma$). On
average, the code attempts one volume change every five
particle-swap moves and 500 displacement moves. Each displacement
move is composed of a simultaneous random translation of the
particle center (uniformly distributed between $\pm 0.05 \sigma$)
and a rotation (with an angle uniformly distributed between $\pm
0.1$ radians) around a random axis.

Standard MonteCarlo simulations (MC) are performed for a system of
$N=$1000 platelets in the NVT ensemble. A MC step is defined as
$N$ attempted moves (defined as in the GEMC). Each state point is
at first equilibrated at $T^*=0.10$, and then quenched down to
$T^*=0.06$, a temperature well below the critical one, where the
system cannot reach equilibrium within the duration of the run.
The waiting time is defined as the time of the quench. To reduce
numerical noise at each waiting time, the observables of interest,
such as the static structure factor $S(Q)=1/N \langle
|\rho_q|^2\rangle$ with $\rho_q=e^{i {\bf Q}\cdot {\bf r}}$, are
averaged over 10 independent runs.

The use of a square-well potential to model the interactions makes
it possible to unambiguously define two platelets as bonded when
the pair-wise interaction energy is  $-u_0$.  Clusters are
identified as groups of bonded platelets.
  To test for percolation, the simulation box is duplicated in all directions, and the ability of the largest
cluster to span the replicated system is controlled. If the
cluster in the simulation box does not connect with its copy in
the duplicated system, then the configuration is assumed to be
nonpercolating. The boundary between a percolating and a
nonpercolating state point is then defined as the probability of
observing infinite clusters in $50\%$ of the configurations.

{\bf Acknowledgments} BR, LZ, and RA
 thank G. Ruocco for his
constant encouragement and advice during the course of this
project. We thank C. De Michele for the code generating the
snapshots of Fig. 2 and ESRF for beamtime. EZ and FS acknowledge
financial support from ERC-226207-PATCHYCOLLOIDS and
ITN-234810-COMPLOIDS.

{\bf Author Contribution} BR, LZ, RA performed experiments. MS,
AM, TN  gave technical support and conceptual advice for the SAXS
experiment . EZ and FS did the modeling and numerical simulations.
All authors discussed the results and implications and contributed
to the writing of the manuscript.

\end{document}